\pdfoutput=1
\documentclass[sigconf]{acmart}

\usepackage{multirow}
\usepackage{tcolorbox}
\usepackage{xcolor}
\usepackage{xspace}
\usepackage[T1]{fontenc}
\usepackage{subcaption}
\usepackage{enumitem}
\usepackage{adjustbox}
\usepackage{balance}

\newcommand{\msmarco}{\textsc{MsMarco}\xspace}
\newcommand{\msmarcovo}{\textsc{Ms Marco-v1}\xspace}

\newcommand{\sift}{\textsc{Sift1M}\xspace}
\newcommand{\deep}{\textsc{Deep1M}\xspace}
\newcommand{\fbssnpp}{\textsc{FB-ssnpp}\xspace}

\newcommand{\kannolo}{\textsc{kANNolo}\xspace}

\newcommand{\svb}{\textsc{Stream VByte}\xspace}

\newcommand{\cocond}{\textsc{CoCondenser}\xspace}
\newcommand{\cocondo}{\textsc{CoCondenser}}
\newcommand{\snowf}{\textsc{SnowFlake}\xspace}
\newcommand{\ebay}{e\textsc{BayProducts7M}\xspace}

\AtBeginDocument{%
  }

\copyrightyear{2026}
\acmYear{2026}
\setcopyright{cc}
\setcctype{by}
\acmConference[CIKM '26]{Proceedings of the 35th ACM International Conference on Information and Knowledge Management}{November 07--11, 2026}{Rome, Italy}
\acmBooktitle{Proceedings of the 35th ACM International Conference on Information and Knowledge Management (CIKM '26), November 07--11, 2026, Rome, Italy}
\acmDOI{10.1145/3799682.3840116}
\acmISBN{979-8-4007-2539-5/2026/11}
\begin{document}

\title{Lossless Compression of Proximity Graph Indexes for Efficient Approximate Nearest Neighbors Search}

\author{Francesco Benocci}
\orcid{0009-0001-5780-8954}
\authornote{All authors contributed equally to this research.}
\affiliation{%
  \institution{ISTI-CNR}
  \city{Pisa}
  \country{Italy}
}
\email{francesco.benocci@isti.cnr.it}

\author{Leonard Dahlmann}
\orcid{0009-0008-5881-8776}
\affiliation{%
  \institution{eBay Inc.}
  \city{Aachen}
  \country{Germany}
}
\email{fdahlmann@ebay.com}

\author{Francesco Gargiulo}
\orcid{0009-0000-5884-4298}
\affiliation{%
  \institution{University of Pisa }
  \city{Pisa}
  \country{Italy}
}
\email{francesco.gargiulo@phd.unipi.it}

\author{Tobias Heuer}
\orcid{0000-0002-5399-0496}
\affiliation{%
  \institution{eBay Inc.}
  \city{Karlsruhe}
  \country{Germany}
}
\email{theuer@ebay.com}

\author{Silvio Martinico}
\orcid{0009-0005-7280-6147}
\affiliation{%
  \institution{University of Pisa \& ISTI-CNR}
  \city{Pisa}
  \country{Italy}
}
\email{silvio.martinico@phd.unipi.it}

\author{Franco Maria Nardini}
\orcid{0000-0003-3183-334X}
\affiliation{%
  \institution{ISTI-CNR}
  \city{Pisa}
  \country{Italy}
}
\email{francomaria.nardini@isti.cnr.it}

\author{Cosimo Rulli}
\orcid{0000-0003-0194-361X}
\affiliation{%
  \institution{ISTI-CNR}
  \city{Pisa}
  \country{Italy}
}
\email{cosimo.rulli@isti.cnr.it}

\author{Rossano Venturini}
\orcid{0000-0002-9830-3936}
\affiliation{%
  \institution{University of Pisa}
  \city{Pisa}
  \country{Italy}
}
\email{rossano.venturini@unipi.it}

\renewcommand{\shortauthors}{Francesco Benocci et al.}

\begin{abstract}
Graph-based Approximate Nearest Neighbor (ANN) indexes, such as HNSW, are the backbone of modern vector search. However, their massive memory footprint creates a severe infrastructure bottleneck.
While vector quantization is widely adopted to reduce the memory required to store data vectors, the graph's edge structure increasingly dominates total memory usage.
Compressing these adjacency lists for online search presents a critical system challenge: because decompression occurs during query processing, i.e., while navigating the graph, memory savings cannot come at the expense of query latency.
In this paper, we present a compression technique explicitly tailored for live HNSW routing. By applying offline graph permutations to artificially induce structural locality, we enable the use of SIMD-accelerated integer compressors, such as \svb, for rapid on-the-fly decompression. Evaluated across diverse real-world workloads, we significantly outperform the current state of the art in both compression ratio and query latency. Our experimental results show that we achieve compression ratios up to $52.7\%$ with negligible penalties in latency compared to the uncompressed baseline. Furthermore, our technique has been effectively deployed in production at eBay, achieving up to $44.3\%$ reduction in edge storage with no degradation in query latency.
 
\end{abstract}




\maketitle

\section{Introduction}
\label{sec:intro}

Modern retrieval systems represent data such as text, images, and products as high-dimensional vector embeddings generated by neural models, including Large Language Models. Retrieval then consists of finding the vectors in a large collection that are most similar to a given query under metrics such as Euclidean distance, dot product, or cosine similarity \cite{ivf_gpu, wang2021comprehensive}.
Since exact nearest neighbor search requires comparing the query against all indexed vectors, its computational cost becomes prohibitive at large scales. Approximate Nearest Neighbor (ANN) methods trade exactness for faster retrieval, and are now a fundamental component of large-scale retrieval systems \cite{aumuller2020ann, wang2021comprehensive}. 
Among the many ANN indexing strategies proposed in the literature, proximity graph methods---most notably Hierarchical Navigable Small World (HNSW)~\cite{malkov2018efficient}---have emerged as one of the dominant approaches due to their strong trade-off between retrieval accuracy and query latency \cite{aumuller2020ann, wang2021comprehensive}. These methods construct a graph over the vectors by connecting nearby points, enabling efficient graph traversal during query processing.
However, they introduce two main sources of memory overhead: storing the vectors and the graph structure. These indexes are typically designed to reside entirely in main memory to guarantee low-latency query processing, making memory consumption a primary bottleneck in large-scale deployments \cite{ottaviano_ann_compression}.
While vector quantization techniques~\cite{pq,lvq,rabitq} are widely adopted for vector compression, the graph structure itself can account for a large fraction of the overall index size, especially when compact vector representations are used \cite{ottaviano_ann_compression}.
As a result, graph compression is particularly relevant for modern ANN systems. At the same time, compressing graph connectivity is particularly challenging in the online in-memory setting, where adjacency lists must be decoded during query execution, and any additional decompression overhead directly affects retrieval latency.
In this line, recent work has begun to explore lossless compression of ANN indexes across a wide spectrum of settings, including online and offline scenarios, inverted file (IVF) indexes \cite{ivf_gpu}, graph indexes, and quantized vector codes \cite{ottaviano_ann_compression}.
However, the current literature still lacks an in-depth exploration of graph compression, which we address here.

In this work,  we provide a thorough exploration of the structural properties of proximity graphs and graph compression techniques for online in-memory ANN search. 
We focus on HNSW~\cite{malkov2018efficient}, one of the most widely adopted graph-based ANN indexes~\cite{aumuller2020ann, survey_graph_2025}. It is available in research libraries such as FAISS~\cite{douze2025faiss}, hnswlib~\cite{hnswlib}, and \kannolo~\cite{delfino2025kann} (that we employ here), and deployed in production by vector database companies including Elasticsearch, Qdrant, Weaviate and Pinecone.

Our contribution revolves around two key components. 
First, we employ offline node permutation techniques~\cite{egb,kaminpar} to improve the compressibility of adjacency lists (Section~\ref{subsec:background_permutations}).
Second, we apply \svb~\cite{streamvbyte} compression, which at query time achieves negligible decompression overhead thanks to its byte-aligned layout and SIMD decompression (Section ~\ref{subsec:graph_comp}).
We evaluate our proposed technique on a broad set of workloads, distinguishing our work from prior studies that present evidence on small, dense vector datasets (\sift~\cite{sift}, \deep~\cite{deep1m}, and \fbssnpp~\cite{fb_ssnpp}). Our benchmarks comprise both classical and neural datasets, over dense and sparse vectors, using Euclidean distance and inner-product similarity, and focus on both compressed and uncompressed vector representations, across datasets ranging from $1$M to $\approx$9M vectors.

The proposed system is currently deployed in production for large-scale ANN retrieval at eBay, whose in-house search engine, Cassini \cite{trotman2017architecture}, uses HNSW for vector retrieval.
We integrated our graph compression framework into Cassini's production pipeline to relieve its memory constraint.
Since its deployment in early 2025 we measure a $44.3\%$ smaller HNSW graph at identical ANN recall, with no measurable latency degradation under live workloads, which lets eBay extend HNSW to more search use cases.

Our main contributions are as follows:
\begin{itemize}[leftmargin=*]
    \item We provide a deep empirical exploration of the HNSW graph structure over three public datasets and a private products dataset from eBay (\ebay) to ground the choice of the adjacency list compression algorithms. Our analysis finds that reference-based compressors such as WebGraph \cite{boldi2004webgraph} are not suitable for compressing ANN graphs, and that node IDs do not reflect the graph's structural topology, highlighting the need to adopt graph permutation techniques \cite{egb,kaminpar} before compressing adjacency lists.
    \item We introduce a compression framework that effectively reduces edge storage while incurring negligible to zero slowdown. We then experimentally show the effectiveness and efficiency of our proposed technique across a wide range of datasets.
    Our approach reduces edge space by up to $52\%$ (\sift) while maintaining query latency within a few percent of the uncompressed baseline and achieving up to $11\%$ lower latency on \ebay at $98\%$ Accuracy@10.
    \item We will release a public implementation of our compression framework within the \kannolo Rust library~\cite{delfino2025kann}.
\end{itemize}

\section{Background and Related Work}
\label{related}

\subsection{ANN Search}

Given a dataset of $N$ vectors in a $D$-dimensional vector space, the $k$-Nearest Neighbors (kNN) problem consists of retrieving the $k$ closest points to a query vector under a given similarity measure \cite{cover1967nearest}. Exact solutions quickly become computationally prohibitive as both $N$ and $D$ grow. In particular, classical data structures such as kd-trees \cite{bentley1975multidimensional} suffer from the curse of dimensionality \cite{indyk1998approximate}, leading to a degradation toward linear scan in high-dimensional settings. 
To overcome these limitations, Approximate Nearest Neighbor (ANN) methods trade small accuracy losses for huge efficiency gains \cite{ferhatosmanoglu2001approximate}, significantly reducing query time and memory access costs \cite{aumuller2020ann}. 

\paragraph{HNSW}
The most effective class of ANN indexes is represented by graph methods~\cite{wang2021comprehensive,aumuller2020ann,survey_graph_2025}, which organize the dataset as a proximity graph by connecting nearby points to facilitate space exploration. The search for a query vector follows a greedy strategy \cite{survey_graph_2025}.

Among graph-based methods, the Hierarchical Navigable Small World (HNSW) graph~\cite{malkov2018efficient} has emerged as one of the most effective and widely-adopted ANN indexes due to its robustness and the strong trade-off between accuracy and efficiency that it offers. 
HNSW organizes vectors into a multi-level proximity graph where higher, sparser levels provide long-range connections to facilitate navigation, while lower levels improve local search accuracy. The lowest level, referred to as the \emph{ground level}, contains all the dataset points. During the search for a query, higher layers are used to quickly locate the query neighborhood, and the ground level is then used to refine the results.

Three standard parameters regulate the effectiveness and efficiency of HNSW: $M$ controls the maximum number of graph connections per node, $\textit{efConstruction}$ regulates the size of the candidate pool during index construction, and $\textit{efSearch}$ controls the exploration breadth during query processing.
In our implementation, query processing is further controlled by a distance-adaptive early-termination threshold $\lambda$ \cite{early-exit}.

\paragraph{Vector Quantization}
ANN indexes require storing the dataset vectors and typically operate in-memory to enable low-latency responses. Therefore, scaling these methods to large, high-dimensional datasets poses a challenge.
To overcome this limitation, much of the literature has focused on \emph{quantization} methods \cite{pq,opq,lvq,rabitq} that reduce the space required for vector data by mapping continuous values to discrete ones.

Product Quantization (PQ) \cite{pq} is one of the most widely adopted quantization techniques \cite{douze2025faiss} in ANN search.
It divides vectors into subspaces, quantizing each independently via $\kappa$-means, achieving compression ratios above 90\% at modest accuracy cost.

\subsection{Graph Compression}
\label{subsec:graph_comp}
In ANN graph indexes, adjacency lists can be arbitrarily ordered. However, by enforcing a sorted order and treating them as monotonically increasing sequences of integers, we can leverage powerful encoding algorithms to significantly reduce memory footprints.

\paragraph{Integer Compression} Elias-Fano~\cite{elias1974} is a quasi-succinct encoding that splits each integer into high and low-order bits: the lower bits are stored explicitly, while the upper ones are stored in a compact unary representation. It compresses sorted adjacency lists and supports direct access through a \textit{Select} operation, which is necessary for decompressing adjacency lists during query routing, since neighborhood lookups are unpredictable and are served on the fly.

Alternatively, a sorted list of integers $S=[x_1,\dots,x_n]$ can be transformed via gap encoding (or delta encoding), yielding $\widetilde{S} = [x_1, x_2-x_1,\dots,x_n-x_{n-1}]$, i.e. by storing the differences between consecutive elements~\cite{gap_encoding, CSUR21}. Ideally, gap encoding shifts the distribution of integers toward much smaller values, reducing the overall entropy of the sequence. 

Entropy coders such as Huffman~\cite{huffman2007method}, Arithmetic coding~\cite{arithmetic1,arithmetic2,arithmetic3} and Asymmetric Numeral Systems (ANS)~\cite{duda2009asymmetric, ottaviano_ann_compression} are near--optimal in space, but due to bit-by-bit decompression introduce unacceptable latency that real-time systems cannot afford. Traditional universal codes (e.g., Elias Gamma or Delta~\cite{elias2003universal}, Fibonacci~\cite{fibonacci}, Zeta~\cite{zeta}) assign fixed, short bit sequences to small integers, yet are not byte-aligned, which slows decoding.
Consequently, modern latency-sensitive compression pipelines typically adopt byte-aligned or word-aligned integer compressors, which sacrifice a small amount of compression ratio in exchange for faster decoding.

Variable Byte (VByte) \cite{witten1999managing} encodes each integer in a variable number of bytes, with continuation bits marking whether more bytes follow. It benefits from gap encoding, since the resulting small integers often fit in a single byte. However, its decoding process involves data-dependent branches and byte-by-byte control flow, limiting instruction-level parallelism and efficient SIMD utilization.

\emph{\svb} (SVB)~\cite{streamvbyte} redesigns the layout of VByte to better match modern hardware. Instead of interleaving control information with the data, it separates the compressed sequence into two streams: a \emph{control stream} and a \emph{data stream}. Each control byte encodes the lengths of multiple consecutive integers; for instance, encoding 32-bit values, a single control byte describes four integers using 2 bits per value, where each 2-bit pattern specifies the number of bytes required to represent the corresponding integer (from 1 to 4 bytes). The data stream then stores the concatenated variable-length representations.
The decoder therefore reads control bytes independently and access the data stream branch-free. With block-level indexing the layout also supports partial decoding and efficient skipping, which fits graph-based ANN search, where only a subset of overall neighbors array is accessed per query.

\paragraph{List-level Compression}
Another dimension along which redundancy can be exploited is at list level, by representing significant portions of one list as a reference to another. Boldi and Vigna observed that this type of redundancy is pervasive in web graphs, where many websites share a large portion of outgoing links~\cite{boldi2004webgraph}. 
They proposed WebGraph, an algorithm that leverages both intra-list and inter-list redundancies to achieve state-of-the-art compression ratios. To mitigate decompression costs, they bound the reference chains by restricting copies to a fixed window of preceding lists. 

While list-level reference compression has become the \textit{de facto} standard for archiving web and social graphs, its applicability to proximity graphs remains largely unexplored. The distinct topological properties of such graphs---combined with the strict latency requirements of their routing algorithms---raise open questions regarding whether exploitable list-level redundancies exist in this domain, and whether the decompression overhead of reference chains can be tolerated during online search.

\paragraph{Previous Work}
A recent work by Severo \emph{et al}.~\cite{ottaviano_ann_compression} from Meta investigates \emph{lossless compression} techniques tailored to ANN data structures, representing the first systematic study of vector identifier compression for ANN indexes.
They show that a substantial fraction of the memory footprint of ANN structures is due to identifiers and graph links rather than vectors, and that these components can be significantly compressed without affecting retrieval accuracy, while highlighting the tension between compression effectiveness and online decoding costs, especially in graph traversal workloads where adjacency lists must be accessed frequently during search.

Their approach rest on the observation that, in both IVF \cite{ivf_gpu} and graph-based indexes, IDs ordering within lists is irrelevant to search correctness.
This invariance enables additional compression opportunities by treating adjacency lists and posting lists either as sorted sequences or directly as sets, allowing the use of succinct representations and entropy coding techniques.
For the online graph scenario \emph{bit-packing} (i.e., storing each identifier of $N$ total elements using $\lceil \log_2 N \rceil$ bits) and \emph{Elias--Fano}~\cite{elias1974} provide the best compression and fast random access.
On IVF indexes, more advanced entropy-oriented approaches, including \emph{Asymmetric Numeral Systems} (ANS)~\cite{duda2009asymmetric, ottaviano_ann_compression}, further exploit the statistical distribution of identifiers and gaps, achieving stronger compression ratios.

\subsection{Graph Permutations}
\label{subsec:background_permutations}
Default ordering of vertices in real-world graphs, dictated by ingestion order, rarely aligns with topology, masking redundancies and leading to suboptimal compression. By finding an opportune bijection of the vertex IDs---a process known as \textit{graph permutation} or reordering---we can artificially induce structural locality and significantly improve the effectiveness of encoding algorithms.

Formally, optimizing for gap encoding can be formulated as the NP-hard \textit{minimum linear arrangement} (MLA) problem, which seeks a node permutation $\phi: V \rightarrow V$ that minimizes the sum of edge lengths, $\sum_{(u,v) \in E} |\phi(u) - \phi(v)|$~\cite{garey1979computers}. Because exact solutions are intractable, modern compression pipelines rely on graph partitioning heuristics. These algorithms cluster the graph into densely-connected regions and assign consecutive IDs to the nodes within each block~\cite{boldi2011layered, slashburn2014, laxman2016compression}.

In the broader graph partitioning literature, two primary paradigms exist, which we leverage to enable effective gap compression: 
\begin{itemize}[leftmargin=*]
    \item \textbf{Recursive Bipartitioning (RB):} Algorithms in this family repeatedly divide the graph into two balanced partitions until the desired number of blocks is reached. The most prominent algorithms in this class include Metis, PaToH, and Scotch~\cite{KarypisK98a,PATOH,PellegriniR96}. Enhanced Graph Bisection (EGB)~\cite{egb} is a specialized variant of RB algorithms that employs a gain-based refinement procedure explicitly designed to optimize the distribution of adjacency-list entries for compression.
    \item \textbf{Direct $k$-way Partitioning:} Alternatively, algorithms in this family partition the graph into $k$ blocks directly. Recent advances have established that direct $k$-way methods generally outperform recursive approaches in both partitioning quality~\cite{KAHYPAR-JOURNAL} and execution speed~\cite{MT-KAHYPAR-JOURNAL}. KaMinPar~\cite{kaminpar}, which employs a highly scalable multilevel scheme, is specifically designed to partition graphs into a large number of blocks (e.g., $k > 1024$), a task where traditional partitioning algorithms often struggle to find feasible solutions or suffer severe performance degradation. This capability is critical for edge compression, as it enables the fine-grained ID grouping required to minimize adjacency list gaps. 
\end{itemize}

By leveraging these partitioning techniques, highly scattered adjacency lists can be transformed into localized sequences with small gaps, unlocking the theoretical bounds of integer compression.

\section{Methodology}
\label{sec:meth}

We structure our methodology into two parts. First, we provide a preliminary analysis of the proximity graph's structural properties.
Second, we describe our compression technique, which builds directly on those findings.

\subsection{Preliminary Analysis}
\label{subsec:pre_analysis}

Consistent with our focus on practical deployments, we restrict our analysis to low-degree HNSW configurations. In our benchmarks, low-degree graphs already achieve high recall; increasing $M$ beyond this regime buys little accuracy while costing memory and, through longer adjacency lists, query latency. Furthermore, HNSW pruning rules keep the edge count well below the cap.
Prior work evaluates degrees from 16 to 256~\cite{ottaviano_ann_compression}. Low-degree graphs are therefore a more challenging and practically relevant target: adjacency lists are shorter, leaving less redundancy to exploit, and the per-list decompression cost is more pronounced relative to the work done per query.

\begin{figure}[b!]
    \centering
    \includegraphics[width=1\linewidth]{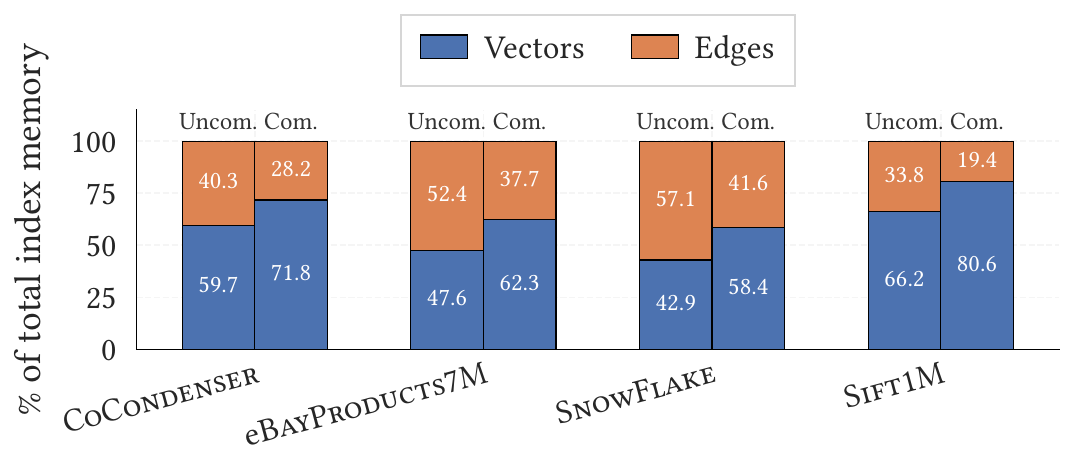}
    \caption{Memory breakdown of HNSW indices across different datasets before (Uncom.) and after (Com.) compression.}
    \label{fig:vectorvsedges}
\end{figure}

\paragraph{Impact of Edges on Index Size}
Proximity graph indexes have two main sources of memory overhead that both scale linearly with dataset size: the stored vectors and the graph edges. 
While vector compression is well-studied and widely deployed, the contribution of edge storage to the overall index size is often underestimated. Figure \ref{fig:vectorvsedges} shows that graph edges constitute a substantial fraction of the total index size, making their compression critical to reducing memory footprint. This is particularly true in quantized settings (Figure \ref{fig:vectorvsedges}, \snowf) and in low-dimensional settings (Figure ~\ref{fig:vectorvsedges}, \ebay), where the dominant cost is given by the graph structure itself rather than the vector data.

\paragraph{Graph Compressibility}
\label{subsubsec:graph_compressibility}
The feasibility of any compression strategy is shaped by the structural properties of the underlying graph. In this section, we examine two such properties that motivate our compression choices: 
\textit{(i)}~the overlap between adjacency lists, which determines whether reference-based compression is profitable; and 
\textit{(ii)}~the distribution of values within adjacency lists, which dictates the choice of integer encoders. 
Table~\ref{tab:dataset-sizes} reports some statistics on node and edge counts of the HNSW graphs under evaluation.

\begin{table}[t!]
\centering
\caption{Node and edge counts and average number of edges per node (E/N) of the ground level of the evaluated HNSW graphs. $^{\star}$ \cocond is made of sparse vectors.
\label{tab:dataset-sizes}}
\begin{tabular*}{\columnwidth}{@{\extracolsep{\fill}}lrrrr@{}}
    \toprule
    Dataset & \multicolumn{1}{c}{Nodes} & \multicolumn{1}{c}{Edges} & \multicolumn{1}{c}{E/N} & \multicolumn{1}{c}{Dim.} \\
    \midrule
    \cocondo$^{\star}$ & 8{,}841{,}823 & 400{,}507{,}947 & 45.3 & $30$,$522$\\
    \ebay & 7{,}262{,}226 & 362{,}099{,}894 & 49.9 & $96$\\
    \sift & 1{,}000{,}000   & 29{,}984{,}015 & 30.0 & $128$\\
    \snowf & 8{,}841{,}823 & 349{,}831{,}589 & 39.6 & $1${,}$024$\\
    \bottomrule
\end{tabular*}
\end{table}

\begin{figure}[b!]
    \centering
    \includegraphics[width=1\linewidth]{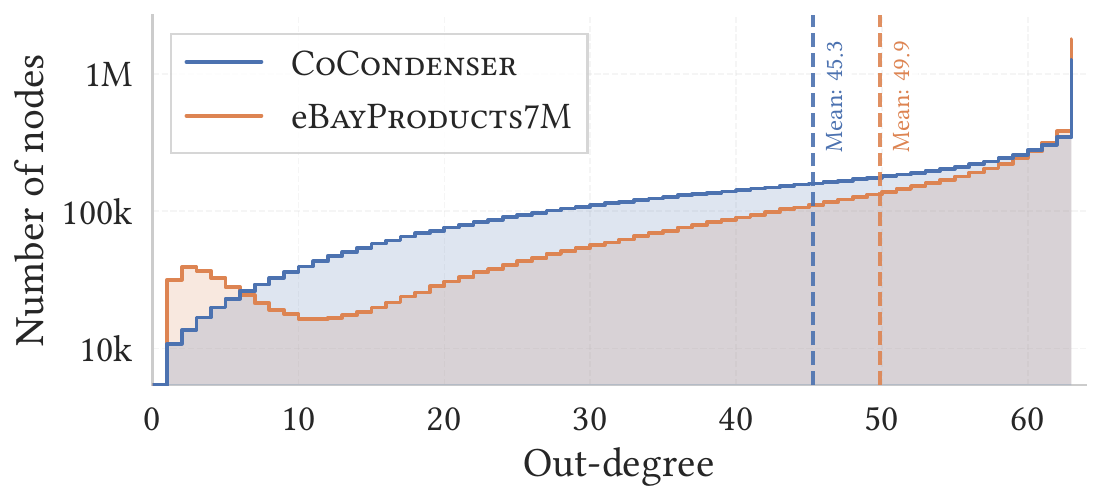}
    \caption{Out-degree distributions of the \cocond and \ebay HNSW graphs (log-scale y-axis). \label{fig:hnsw-out-degree-distribution}}
\end{figure}

Both properties rest on one structural constraint: HNSW enforces a strict degree bound on every node to balance routing efficiency, memory consumption, and graph navigability. This sharply contrasts with web and social networks, whose power-law degree distributions produce high-degree hubs that reference-based compressors exploit for list overlap. As shown in Figure~\ref{fig:hnsw-out-degree-distribution}, nodes in \cocond and \ebay concentrate near this cap of 64 edges per node, with a substantial fraction of nodes saturating it entirely. These short, bounded adjacency lists yield two immediate consequences: first, the absolute maximum number of edges that any two lists can share is intrinsically small; second, any per-list metadata overhead must be amortized over only a few dozen values.

\paragraph{List-level Compression}
To evaluate whether these approaches are viable for HNSW graphs, we compute an ideal upper bound on the savings achievable via reference copying.

Let $N(x)$ denote the out-neighborhood of $x$, and let $k_{u,v} = |N(u) \cap N(v)|$. Encoding $u$ as a copy from a reference $v$ requires identifying which $k_{u,v}$ elements of $N(v)$ are reused, which costs at least $\log_2 \binom{|N(v)|}{k_{u,v}}$ bits under a uniform distribution over subsets. While a larger $|N(v)|$ increases the maximum number of copyable edges, it also raises the combinatorial selection cost.

For each node $u$, let $S(u)$ denote the maximum savings, in bits, attainable by encoding $u$'s adjacency list as a partial copy of a reference node. Taking bit-packing as the per-edge cost $B$ and exhaustively searching over all candidate references yields:

\begin{equation}
    S(u) = \max_{v\neq u}\left(k_{u,v}\cdot B - \log_2 \binom{|N(v)|}{k_{u,v}}\right).
    \label{eq:best-reference-bits-savings}
\end{equation}

The first term counts the bits saved by the $k_{u,v}$ copied edges; the second is the selection cost defined above.

Equation~\ref{eq:best-reference-bits-savings} is deliberately optimistic: it allows cyclic and unbounded reference chains, ignores the cost of encoding the reference ID itself, and presupposes an ideal node ordering in which the optimal $v$ always lies within $u$'s reference window. Each of these assumptions would be violated under any practical implementation, so $S(u)$ strictly upper-bounds the achievable savings.

On \ebay, although a profitable reference exists for over 99.9\% of nodes, the aggregate ceiling on space savings is just 8.8\%. The best reference covers only a small fraction of each list: median 8.1\%, 90th percentile 22.7\%. The average list for a node $u$ contains 49.9 edges (1147 bits) and shares only $k_{u,v} = 5.2$ edges with its optimal reference $v$, yielding savings of $k_{u,v}\cdot B \approx 120$ bits. The selection cost is $\log_2 \binom{|N(v)|}{k_{u,v}}\approx 19$ bits, for a net reduction of just $S(u) = 101$ bits per list. Results for \cocond are nearly identical and omitted for brevity.

Given that this ideal ceiling is already low and that any practical implementation would incur an additional latency penalty for chasing reference chains during random access, we conclude that reference-based compression offers an unfavorable trade-off for HNSW graphs and exclude it from further evaluation.

\paragraph{Value Distribution and Reordering}
\label{par:entropy-gaps}
Having ruled out reference-based encoding, we turn to the second property: the distribution of values within adjacency lists. We characterize this distribution through its zero-order empirical entropy $H_0(X) = -\sum_i p_i \log_2 p_i$, computed globally over all nodes of the graph.

\begin{figure}[t!]
    \centering
    \includegraphics[width=1\linewidth]{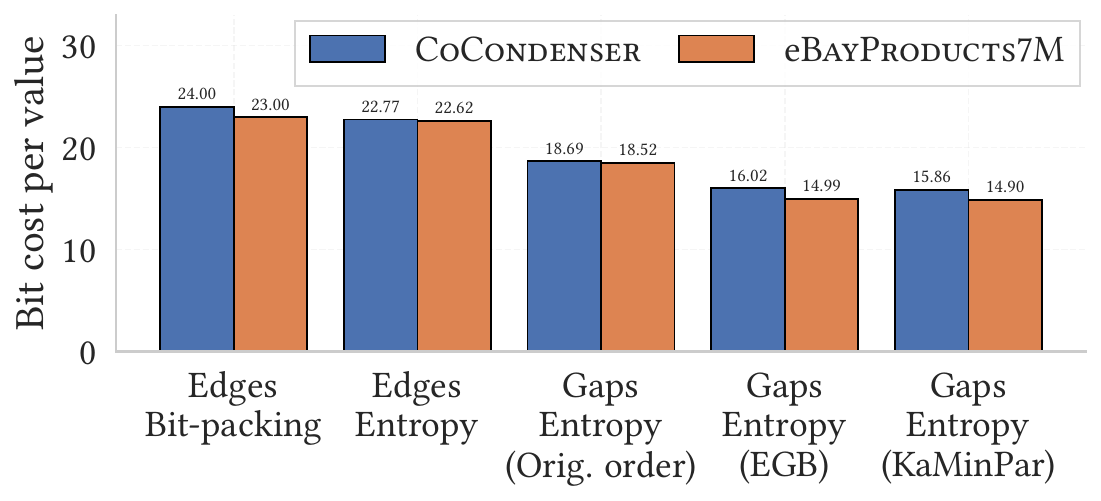}
    \caption{Average bit per value for edges and gaps.}
    \label{fig:reorder-entropy}
\end{figure}

The adjacency lists of HNSW graphs exhibit high zero-order entropy: as shown in Figure~\ref{fig:reorder-entropy}, the per-value cost of bit-packing exceeds $H_0$ by at most $1.23$ bits. This indicates that the distribution of edge endpoints is close to uniform across the ID space, making fixed-width packing nearly optimal.

The standard remedy is gap encoding, which concentrates probability mass on small values and thereby lowers $H_0$. Applied directly, it reduces the per-value entropy from 22.77 to 18.69 bits on \cocond and from 22.62 to 18.52 bits on \ebay---a meaningful but insufficient reduction, since the gaps still span a large fraction of the ID space. 

\begin{figure}[b]
    \centering

        \includegraphics[width=\linewidth]{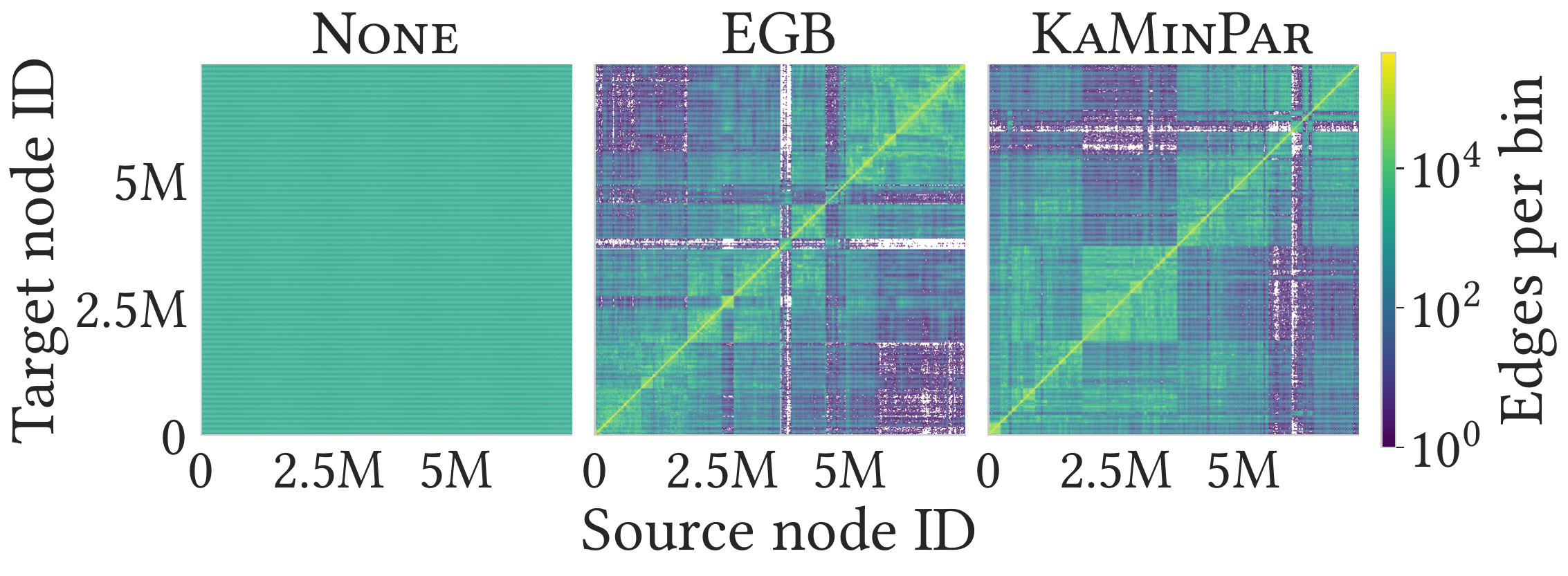} 
    
    \caption{Log-scaled edge density in the \ebay graph. Before permutation (left), edges are spread uniformly across the ID space. Applying permutations heavily reshapes this distribution by concentrating the edge mass along the diagonal. We observe similar patterns on \cocond.}
    \label{fig:edge-locality-reordering}
\end{figure}

The remaining reduction must come from the ID assignment itself: nodes that frequently co-occur in adjacency lists should receive numerically close IDs, so that intra-list gaps concentrate near zero. The randomness of the default assignment (None) is visible in Figure~\ref{fig:edge-locality-reordering}, where the edge density is spread uniformly across the ID space. Applying the two offline reordering algorithms introduced in Section~\ref{subsec:background_permutations}---EGB and KaMinPar---produces a profound concentration of edge mass along the diagonal, confirming that adjacent IDs increasingly correspond to graph-adjacent nodes. The impact on $H_0$ is substantial (Figure~\ref{fig:reorder-entropy}): after EGB, the gap entropy drops to 16.02 bits on \cocond and 14.99 bits on \ebay; KaMinPar tightens these to 15.86 and 14.90 bits, respectively. These post-reordering entropies define the lower bound that the integer codecs evaluated in the next section must approach.

Beyond compression, graph reordering increases vectors spatial locality, making cache reuse and hardware prefetching more effective.
Coleman \emph{et al}.~\cite{ann_graph_permutation} show that this can reduce cache and TLB misses and improve HNSW query latency by up to 40\%.
We observe the same trend in our setting, with reordering reducing the average query time of the uncompressed baseline from 3.0\% (\cocond) to 17.0\% (\ebay).
However, since final ANN results must be reported using the original dataset IDs, when reordering we must store the inverse permutation, thus introducing additional memory overhead. For this reason, we enable it only when the compression method benefit from it.
When enabled, we compact this mapping with fixed-width bit packing, using $\lceil \log_2 N \rceil$ bits per identifier for a dataset of $N$ elements.

Given the small difference ($<1\%$) between EGB and KaMinPar in this context (Figure \ref{fig:reorder-entropy}), we adopt EGB as the default permutation strategy. This decision is driven by two key factors. Theoretically, EGB's gain-based refinement is explicitly tailored to minimize intra-list gaps for compression, whereas KaMinPar optimizes a more general $k$-way edge-cut objective. From a systems engineering perspective, EGB provides a lightweight, native Rust implementation that integrates seamlessly into our Rust-based indexing pipeline.

\paragraph{Integer Compression Candidates}
Having established that offline graph permutations coupled with gap encoding significantly reduce the entropy of adjacency lists, the remaining challenge is selecting an integer compressor that can approach this theoretical limit without introducing prohibitive query-time decoding latency. 

\begin{figure}[b!]
    \centering
    \includegraphics[width=1\linewidth]{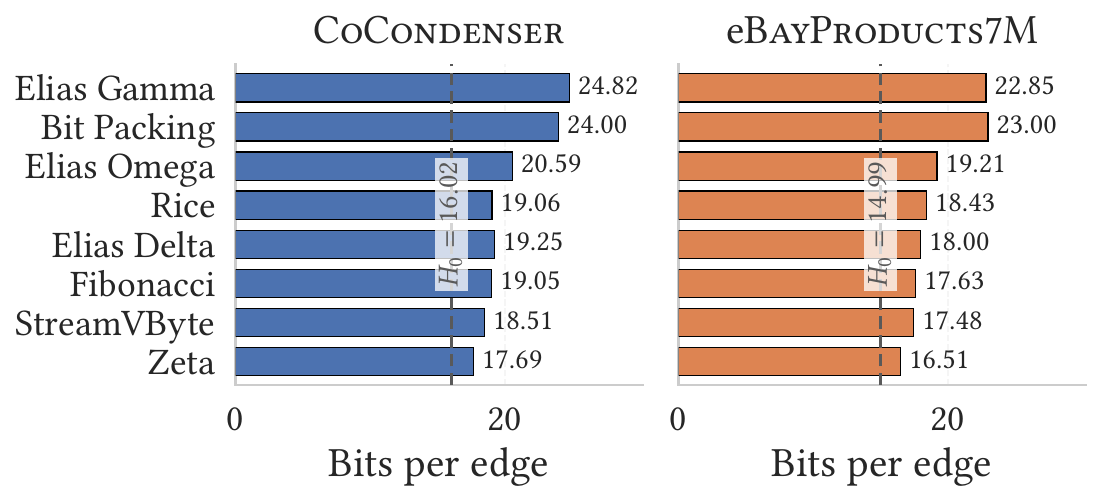}
    \caption{Average bits per edge achieved by various integer compression algorithms on EGB-permuted gap lists. The dashed line marks the zero-order entropy $H_0$. For parametric algorithms we report the best configuration.}
    \label{fig:bits-edge-analysis-entropy}
\end{figure}

Figure~\ref{fig:bits-edge-analysis-entropy} surveys integer compression algorithms.
Among universal codes, Zeta~\cite{zeta} achieves the smallest practical footprint. \svb~\cite{streamvbyte} sacrifices less than 1 bit/edge relative to Zeta, enabling branch-free SIMD decoding. The remaining codecs occupy unfavorable positions on both axes and are not considered further.

\paragraph{Auxiliary Structures}
\kannolo employs a variable-degree representation for adjacency lists, where neighbors are stored contiguously in a flat array, and an auxiliary offset vector \texttt{off} identifies the boundaries of each neighborhood. In particular, the neighbors of node $i$ are stored in positions $\texttt{off}[i], \dots, \texttt{off}[i+1]$ of the array.

Since offsets form a monotone sequence whose gaps correspond to adjacency-list lengths, we compress them using Elias--Fano encoding~\cite{elias1974}, which stores a monotone sequence of $n$ integers drawn from a universe of size $U$ in $n\!\left(2 + \log_2 \frac{U}{n}\right) + o(n)$ bits. In \texttt{off}, the ratio $U/n$ corresponds to the average gap between consecutive offsets, which is $U/n \leq 2M$, yielding $2 + \log_2 \frac{U}{n} \leq 2 + \log_2(2M)$. 
In our setting ($2M=64$), this translates into barely $8 + \frac{o(n)}{n}$ bits per element, a large reduction from the $64$ bits per element of the original array. Furthermore, Elias-Fano allows us to efficiently perform random accesses and decompress single values of \texttt{off}, which is exactly what we need during query processing.

The analysis above fixes our pipeline. 
First, we apply the EGB algorithm to permute the node IDs, storing the compacted permutation mapping. Second, we gap-encode the adjacency lists, compress them using \svb and compress the index's offset array using Elias--Fano to guarantee maximum overall memory efficiency.

 \subsection{Implementing our Compression Framework}
\label{subsec:system_implementation}
We implement our compression framework, alongside all evaluated baselines, within the HNSW index of the \kannolo~\cite{delfino2025kann} library. \kannolo represents a highly optimized, uncompressed baseline, thus ensuring that any measured compression overhead is genuine and not masked by inefficiencies in the underlying implementation.

Building upon \kannolo's architecture, we implement an abstraction layer for adjacency lists that isolates their physical representation from the logical graph traversal. Consequently, the core HNSW search algorithm remains entirely unmodified at query time, allowing various compressed (and uncompressed) backends to be plugged in seamlessly behind the same unified interface.

Under this abstraction, gap-based codecs such as \svb and Zeta apply compression at the granularity of individual adjacency lists rather than across a flattened global array. By aligning decoding entry points with list boundaries, each neighborhood can be decompressed independently on demand. This perfectly matches the random-access node expansion pattern of HNSW search, avoiding the overhead of scanning unrelated neighborhoods. To further minimize CPU overhead during ground-level expansion, decompressed neighbors are buffered into small, stack-allocated chunks, eliminating temporary heap allocations and repeated decoding.

Finally, when graph reordering is enabled, traversal operates natively within the reordered local identifier space. Because the dataset vectors are permuted along with the graph node IDs offline, the search's critical path requires no ID translation. Only the search results are translated back to the original dataset IDs via the stored mapping before being returned to the user.

\section{Experiments}
\label{sec:exp}

\begin{table*}[h!]
\centering
\caption{Average query time ($\mu$s) at different Accuracy@10 cut-offs (90, 94, 98). The space for each compression method is reported as bits/edge (b/e), with the compression ratio over Plain in gray.}
\label{tab:all_datasets}

\begin{tabular}{@{} l @{\hspace{1em}} *{4}{c} c *{4}{c} c *{4}{c} c *{4}{c} @{}}
\toprule
& \multicolumn{4}{c}{\textbf{\cocond}} & & \multicolumn{4}{c}{\textbf{\ebay}} & & \multicolumn{4}{c}{\textbf{\sift}} & & \multicolumn{4}{c}{\textbf{\snowf} (PQ)} \\
\cmidrule(lr){2-5} \cmidrule(lr){7-10} \cmidrule(lr){12-15} \cmidrule(lr){17-20}
& b/e & 90 & 94 & 98 & & b/e & 90 & 94 & 98 & & b/e & 90 & 94 & 98 & & b/e & 90 & 94 & 98 \\
\midrule
Plain                & 33.4 {\scriptsize\textcolor{gray}{\phantom{1.0$\times$}}} & 311 & 386 & 778 & 
                     & 33.3 {\scriptsize\textcolor{gray}{\phantom{1.0$\times$}}} & 185 & 264 & 632 & 
                     & 34.1 {\scriptsize\textcolor{gray}{\phantom{1.0$\times$}}} & \textbf{55} & \textbf{69} & \textbf{111} &
                     & 33.6 {\scriptsize\textcolor{gray}{\phantom{1.0$\times$}}} & 261 & 302 & 497 \\
Bitpacker            & 24.1 {\scriptsize\textcolor{gray}{1.3$\times$}} & 315 & 391 & 791 & 
                     & 23.1 {\scriptsize\textcolor{gray}{1.4$\times$}} & 193 & 277 & 668 & 
                     & 20.2 {\scriptsize\textcolor{gray}{1.6$\times$}} & 57 & 71 & 115 &
                     & 24.1 {\scriptsize\textcolor{gray}{1.3$\times$}} & 266 & 310 & 509 \\
Elias-Fano           & 23.6 {\scriptsize\textcolor{gray}{1.4$\times$}} & 332 & 415 & 843 & 
                     & 23.2 {\scriptsize\textcolor{gray}{1.4$\times$}} & 231 & 333 & 832 &  
                     & 20.7 {\scriptsize\textcolor{gray}{1.6$\times$}} & 79 & 89 & 137 &
                     & 23.9 {\scriptsize\textcolor{gray}{1.4$\times$}} & 288 & 338 & 563 \\
Zeta                 & 18.7 {\scriptsize\textcolor{gray}{1.7$\times$}} & 327 & 406 & 819 & 
                     & 17.4 {\scriptsize\textcolor{gray}{1.9$\times$}} & 214  & 302  & 714 & 
                     & 14.6 {\scriptsize\textcolor{gray}{2.3$\times$}} & 70 & 87 & 141 &
                     & 17.0 {\scriptsize\textcolor{gray}{1.9$\times$}} & 277 & 324 & 545 \\
\textbf{SVB}           & 19.4 {\scriptsize\textcolor{gray}{1.7$\times$}} & \textbf{309} & \textbf{384}  & \textbf{765}  & 
                     & 18.3 {\scriptsize\textcolor{gray}{1.8$\times$}} & \textbf{177} & \textbf{250}  & \textbf{564}  &  
                     & 16.1 {\scriptsize\textcolor{gray}{2.1$\times$}} & 58  & 71 & 114  &
                     & 18.0 {\scriptsize\textcolor{gray}{1.8$\times$}} & \textbf{260} & \textbf{302} & \textbf{494} \\
\bottomrule
\end{tabular}
\end{table*}

\smallskip \noindent
We conduct our experiments on four datasets spanning a diverse range of settings: image descriptors, text passages encoded with both dense and sparse models, and product embeddings (Table~\ref{tab:dataset-sizes}).
\begin{itemize}[leftmargin=*]
    \item \sift \cite{sift} is arguably the most adopted dataset for ANN benchmarking~\cite{aumuller2020ann}, consisting of $1$M $128$-dimensional image descriptors produced using the Scale-Invariant Feature Transform (SIFT)~\cite{sift} algorithm. It is evaluated with Euclidean distance on $10$K queries.

    \item \msmarcovo~\cite{msmarco} is a collection of $\approx\!8.8$M text passages widely adopted in Information Retrieval benchmarks, with a query set of $6{,}980$ queries (\texttt{dev.small}). We produce two different embedding sets and evaluate both under inner product similarity.
    \begin{itemize}[leftmargin=*]
    
    \item \snowf~\cite{snowflake} is a lightweight state-of-the-art dense encoder producing $1024$-dimensional vectors. As its encoding of \msmarco exceeds $30$ GB, we compress it via Product Quantization (PQ) with $M=128$ subspaces, achieving a $32\times$ reduction in storage. Since heavy compression degrades accuracy, we adopt a two-stage search: the top-$k^{\prime}$ ($k^{\prime} > k$) candidates are retrieved using PQ-encoded vectors, then reranked with the uncompressed representations to select the final top-$k$. This mimics the practical scenario in which the reranking step accesses uncompressed data from disk, while the compressed vectors and the graph structure are kept in main memory.
    
    \item \cocond~\cite{cocondenser} is a sparse encoder built on the BERT~\cite{bert} vocabulary, resulting in output vectors that live in an $\approx\!30$K-dimensional space. The vectors have $120$ non-zero components on average, while queries have $42$.  We employ the \texttt{dotvbyte} algorithm~\cite{bruch2026forward} to store the components and $8$ bits floating point values to store the values.
    
    \end{itemize}

    \item \ebay: this is a set of roughly $7$M $96$-dimensional product embeddings from eBay. The embedding model was fine-tuned for product search from an in-house variant of BERT specialized for the e-commerce domain. The query set consists of $10$K query embeddings.
\end{itemize}

\paragraph{Metrics} We evaluate edge-compression methods along three dimensions: memory footprint, search effectiveness, and search efficiency.
Memory footprint is measured as the average number of bits per edge; it accounts for the neighbor lists (on all HNSW levels), the offsets array \texttt{off}, and, optionally, the reverse mapping of the permutation.
Search effectiveness is measured as Accuracy@10, defined as the fraction of the true 10 nearest neighbors among the top-10 retrieved results, averaged over all queries.
Search efficiency is measured as the average query latency in microseconds ($\mu$s). For each target Accuracy@10 level in $\{90, 94, 98\}$, we report the latency of the fastest configuration that reaches that level.

\paragraph{Hyperparameters} We build all HNSW graphs with $M=32$ and set \textit{efConstruction} to $200$ for \sift, $500$ for \snowf, $1{,}000$ for \ebay, and $2{,}000$ for \cocond. We sweep \textit{efSearch} in $[10,19]$ and $\lambda$ in $[0.03,0.15]$. The selected (\textit{efSearch}, $\lambda$) pairs at $90\%$, $94\%$, and $98\%$ Accuracy@10 are: $(11,0.06)$, $(11,0.10)$, $(15,0.12)$ for \sift; $(10,0.03)$, $(12,0.05)$, $(16,0.12)$ for \cocond; and $(10,0.10)$, $(12,0.12)$, $(19,0.15)$ for \ebay.
For \snowf, we make both \textit{efSearch} and $k^{\prime}$ vary in $[20, 100]$; we deactivate early exit to avoid blowing up the parameter space. The best results for this dataset are obtained with (\textit{efSearch}; $k^{\prime}$) pairs respectively $(25,20)$, $(35,30)$ and $(80,80)$.

For the Zeta codec, $\kappa$ controls the trade-off between codeword length and decoding efficiency. We set $\kappa=5$ (\ebay, \cocond) or $\kappa=4$ (\sift, \snowf) as these values maximize compression without degrading query efficiency.

\paragraph{Hardware Platform} Our experiments run on an Intel Xeon Silver 4314 CPU @ 2.40GHz with 32 physical threads and 32 hyperthreaded ones. Our implementation is written in Rust \texttt{1.95.0}-\texttt{nightly}, integrated into the \kannolo~\cite{delfino2025kann} library, which also supports the RerankIndex for the experiments with \snowf~\cite{martinico2026multivector}.

\paragraph{Results}
Table~\ref{tab:all_datasets} reports our evaluation across the four datasets. \emph{Plain} represents the uncompressed baseline.
Bitpacker provides a lightweight compromise, reducing memory usage by $1.3\times$--$1.6\times$ while keeping latency close to the uncompressed baseline across all datasets.
Elias--Fano is dominated on both axes: it is consistently the slowest method while compressing no better than Bitpacker.
Zeta achieves the smallest memory footprint across all datasets, with reductions from $1.7\times$ (\cocond) to $2.3\times$ (\sift) compared to Plain. However, we pay this compactness with slow downs up to $\approx27\%$ on \sift and $\approx14\%$ on \ebay.
\svb yields the best operating point across all four benchmarks, combining strong compression with efficient decoding. Its memory footprint remains close to that of Zeta, within roughly $4$--$10\%$, while its query latency matches or even improves over Plain on \cocond, \ebay, and \snowf, with only a marginal overhead on \sift. \svb is therefore the only method that simultaneously approaches the compression ratio of Zeta and the query performance of the uncompressed baseline.

Considering the graph permutation as a fraction of the total build time (graph construction plus permutation), reordering accounts for just 3.5\% on \cocond, 7.8\% on \snowf, 15\% on \ebay, and 30\% on \sift.
Permutation time depends on the number of edges and their distribution, while graph construction time depends on \textit{efConstruction} and on vector dimensionality. \sift is where this fraction peaks: low dimensionality and no need for a high \textit{efConstruction}.
Because index construction is a one-time, offline procedure, permutation's overhead represents a highly favorable trade-off for the memory reductions and latency improvements demonstrated by our system. 

\section{Conclusions and Future Work}
\label{sec:conclusion}

In this work, we study the problem of compressing the adjacency lists of proximity graphs for ANN search. An offline node permutation induces structural locality in adjacency lists, substantially lowering their entropy under gap encoding. We then compress the reordered, delta-encoded neighbor sequences with \svb, organized into adjacency-list-aligned blocks for fast SIMD decoding at query time.
We evaluate our framework on four diverse benchmarks spanning images, dense and sparse text embeddings, and product embeddings. Across all datasets, \svb provides the best operating point, with a compression ratio up to $2.1\times$ with zero to negligible penalties in latency compared to the uncompressed baseline. Beyond public benchmarks, the framework has been deployed in production at eBay, reducing HNSW graph memory by $44.3\%$ with no measurable degradation in query latency.

As per future work, we plan to extend our framework beyond HNSW to other graph-based ANN indexes such as Vamana~\cite{vamana} or NSG~\cite{fu2017fast}, assessing how the compression-latency trade-off shifts across different graph structures. A second direction is to co-design graph and vector compression in a unified, compression-aware index, jointly optimizing edge encoding and vector quantization to unlock further memory savings, particularly in regimes where both components contribute comparably to the overall index size.

\section*{GenAI Usage Disclosure}
During the preparation of this work, the authors used Gemini for grammar and spelling checking, paraphrasing, and rewording of existing text.
Claude Sonnet 4.6 was used to assist in writing and debugging code.
After using these tools, the authors reviewed and edited the content as needed and take full responsibility for the publication's content.

\bibliographystyle{ACM-Reference-Format}
\balance
\bibliography{sample-base.bib}

\end{document}